\documentclass[10pt]{article}

\usepackage{amssymb,amsbsy}
\usepackage{amsmath}
\usepackage{amsthm}
\usepackage{graphics,graphicx}
\usepackage[T1]{fontenc}
\usepackage{color}
\usepackage{subfigure}

\newcommand{\beq}{\begin{equation}}
\newcommand{\eeq}{\end{equation}}

\newcommand{\beqar}{\begin{eqnarray}}
\newcommand{\eeqar}{\end{eqnarray}}
\newcommand{\bit}{\begin{itemize}}
\newcommand{\eit}{\end{itemize}}
\newcommand{\benum}{\begin{enumerate}}
\newcommand{\eenum}{\end{enumerate}}
\newcommand{\barr}{\begin{array}}
\newcommand{\earr}{\end{array}}
\def\ds{\displaystyle}

\newcommand{\modIII}{\text{III}}

\def\XXint#1#2#3{{\setbox0=\hbox{$#1{#2#3}{\int}$}
   \vcenter{\hbox{$#2#3$}}\kern-.5\wd0}}

\def\b0{\mbox{\boldmath $0$}}

\def\bc{\mbox{\boldmath $c$}}

\def\bx{\mbox{\boldmath $x$}}

\def\bB{\mbox{\boldmath $B$}}

\def\bY{\mbox{\boldmath $Y$}}

\def\f0{\ensuremath{\mathbb{O}}}

\newcommand{\bmM}{\mbox{\boldmath $\mathcal{M}$}}

\title{Interaction of an interfacial crack with linear micro-defects under out-of-plane shear loading}

\author{G. Mishuris$^{(1)}$, A. Movchan$^{(2)}$, N. Movchan$^{(2)}$, A. Piccolroaz$^{(1)}$, 
\\
\\
$^{(1)}$
{\it Institute of Mathematical and Physical Sciences, Aberystwyth University,} \\ 
{\it Ceredigion SY23 3BZ, Wales, U.K.} \\
$^{(2)}$
{\it Department of Mathematical Sciences, University of Liverpool,} \\
{\it  Liverpool L69 3BX, U.K.} \\
}

\begin{document}

\maketitle

\begin{abstract}
\noindent
The interaction of an interfacial crack with small impurities is analysed on the basis of an asymptotic formula derived by the authors.
The interaction between the main crack and the defects (e.g. small cracks or inclusions) is described asymptotically by analysing the dipole 
fields and the corresponding dipole matrices of the defects in question. The method is generic, 
and  it serves interfacial cracks with general distributed loading on the crack faces, taking into account possible asymmetry in the boundary 
conditions, and in a particular configuration for a crack in a homogeneous medium results agree with those obtained earlier by \cite{gong-1995}. 
Shielding and amplification effects of the defects on the propagation of the main crack along the interface are investigated. Numerical 
computations based on the explicit analytical formulae show potential applications in the design of composite and fiber reinforced materials.
\end{abstract}


\newpage

\tableofcontents
  
\newpage

\section{Introduction}
\label{sec01}

Asymptotic models of a brittle crack interacting with a small defect (or a finite number of micro-defects) have been addressed in 
(\cite{atkinson-1972,sendeckyj-1974,rubinstein-1986,rose-1986,gong-1989,gong-1992,movchan-1995,valentini-jam-1999}). A finite crack 
interacting with a large number of micro-defects in an elastic solid was also considered in (\cite{kachanov-1987,movchan-2002}); a survey 
of publications on macro-microcrack interaction problems is included in \cite{petrova-2000}.  The approach \cite{movchan-2002} has led to 
homogenization approximations for dilute composite media. Efficient numerical algorithms for cracks in solids have been developed for 
formulations based on singular integral equations. In particular, numerical techniques have been adopted 
to solve macrocrack--microdefects interaction problems (\cite{romalis-1984,tamuzs-1999,kanaun-2003}). The papers 
(\cite{ortiz-1987,hutchinson-1987}) led to the notion of shielding/amplification due to the presence of micro-defects. 
A formal iterative procedure developed for an integral equation formulation describing interaction between a finite size interfacial crack 
and near-interface smaller cracks is discussed in \cite{Wang1,Wang2}; the solutions of related integral equation involved the Chebyshev 
polynomial representations. Interaction of a dislocation dipole with a Mode III interfacial crack was studied in \cite{Bo}; the results 
included the analysis of the ``shielding effect'' for the case when the dislocation dipole is placed in a neighbourhood of the crack tip. 
The classical mathematical model for bi-material media containing cracks propagating along the interface between different phases has been developed by 
\cite{willis-1971a}.

\begin{figure}
\begin{center}
\includegraphics[width=12cm]{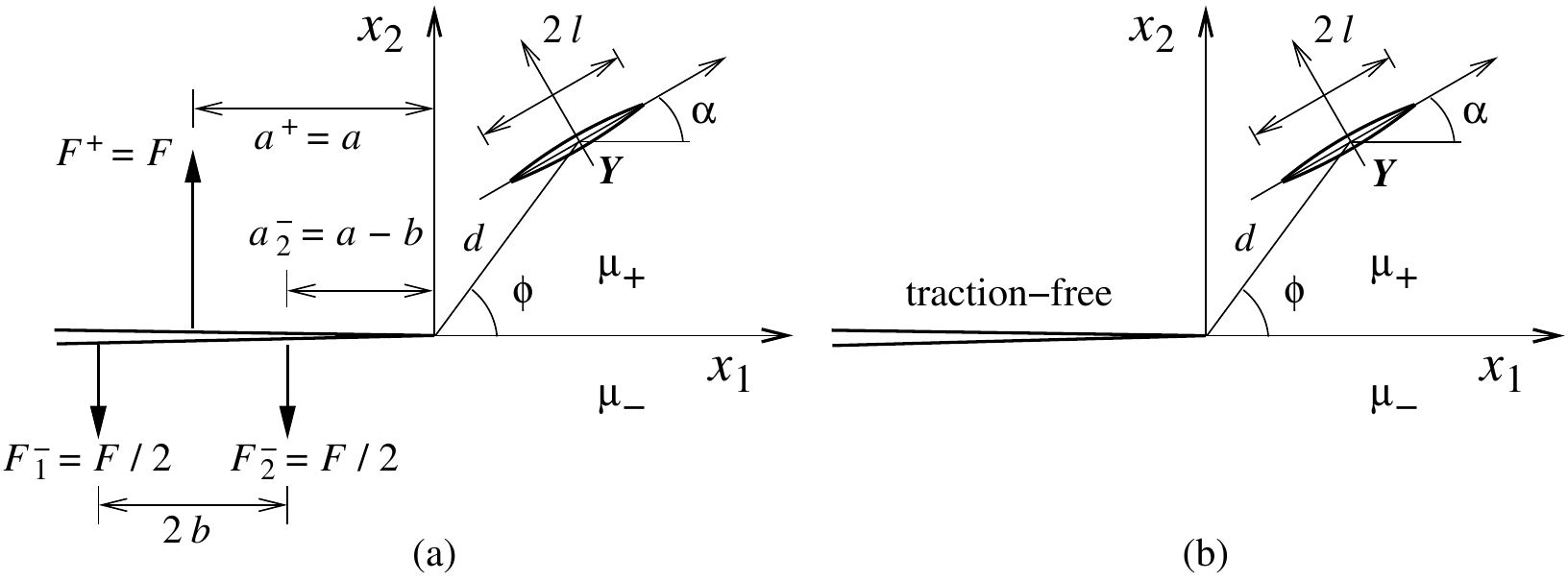}
\caption{\footnotesize A line defect (microcrack/rigid line inclusion) of length $2l$ arbitrarily oriented and situated at a distance $d$ 
from the tip of the main interfacial crack. Fig.\ \ref{fig01}(a) corresponds to the problem under consideration. Fig.\ \ref{fig01}(b) 
illustrates a configuration where the remote load is characterised via the value of the stress intensity factor.}
\label{fig01}
\end{center}
\end{figure}

In this paper we consider structures consisting of two different elastic materials, with the geometry and main notations shown in Fig.\ \ref{fig01}. 
It is assumed that the middle point $\bY$ of the micro-defect is situated at a distance $d$ from the tip of the main interfacial crack and the 
micro-defect length $2l$ is much smaller compared to $d$ (that is, $\varepsilon = l/d$ is a small parameter,
$0 < \varepsilon \ll 1$).

Our asymptotic procedure can be summarised as follows. As $\varepsilon \to 0$, the asymptotic solution is sought in the form
\begin{equation}
\label{dipolar}
u(\bx) = v^{(0)}(\bx) + \varepsilon W^{(1)} \left(\frac{\bx - \bY}{\varepsilon}\right) +
\varepsilon^2 v^{(2)}(\bx) + O\left(\varepsilon^3\right),
\end{equation}
where $\bx = (x_1,x_2)$. The first term $v^{(0)}$ corresponds to an unperturbed problem for a single interfacial crack. The second term 
$W^{(1)}$ denotes the boundary layer appearing near the small defect. The third term $v^{(2)}$ is the solution of the problem for the 
unperturbed solid with the interfacial crack loaded along the crack surfaces by tractions induced by the boundary layer $W^{(1)}$.

For a particular case of a crack in a uniform elastic body and the remote load applied away from the crack tip, our asymptotic approximation 
agrees with that of \cite{gong-1995}:
\begin{equation}
\label{gong_1995} 
\frac{\Delta K_\modIII}{K_\modIII} =
\frac{\varepsilon^2}{4} \cos\left(\frac{3\phi}{2} - \alpha\right) 
\cos\left(\frac{\phi}{2} - \alpha\right) + O\left(\varepsilon^4\right).
\end{equation}

The procedure outlined above is the essence of the so-called general dipole matrix approach developed in \cite{movchan-2002} and described 
for various types of small defects in \cite{roaz-2011}. Here, we employ the representation for an interfacial crack developed in 
\cite{roaz-jmps-2009} where the subscripts $\pm$ correspond to the position of the point under consideration within dissimilar half-planes 
(see Fig.\ \ref{fig01}). 

The present paper addresses the accuracy of the proposed asymptotic approximation and it analyses the discrepancy in the traction transmission 
conditions on the interface due to the boundary layer $W^{(1)}$ and hence its contribution to the perturbation of the stress intensity factor.

\section{Preliminary results (unperturbed crack)}
\label{sec02}

Using the weight function approach, the stress intensity factor for the interfacial crack loaded by given out-of-plane tractions $p^\pm(x_1)$ 
($x_1 < 0$) can be evaluated as 
\beq
\label{SIF_1} 
K_\modIII^{(0)} = - \sqrt{\frac{2}{\pi}}
\int_{-\infty}^0 \frac{\mu_- p^+(x_1) + \mu_+ p^-(x_1)}{\mu_+ + \mu_-} (-x_1)^{-1/2} dx_1.
\eeq

Without loss of generality, we formulate the results in terms of concentrated forces applied on the crack surfaces, so that the formula 
(\ref{SIF_1}) can be rewritten in the following form:
\beq
\label{SIF_2}
K_\modIII^{(0)} = -\frac{1 + \eta}{\sqrt{2\pi}} \sum_{j=1}^{n} F^+_j \big(a^+_j\big)^{-1/2} -
\frac{1 - \eta}{\sqrt{2\pi}} \sum_{j=1}^{m} F^-_j \big(a^-_j\big)^{-1/2}.
\eeq
Here $\eta$ is the contrast parameter, $\eta = (\mu_- - \mu_+)/(\mu_- + \mu_+),$ and the point forces $F_j^+$ are applied at the points 
$(-a^+_j,0+)$, ($j=1,...,n$), while $F_j^-$ are applied at the points $(-a^-_j,0-)$, ($j=1,...,m$).

The formula (\ref{SIF_2}) can then be written in the form
\beq
\label{SIF_3}
K_\modIII^{(0)} = \sum_{j=1}^{n} K_{III,j}^{(0,+)} + \sum_{j=1}^{m} K_{III,j}^{(0,-)}.
\eeq

We note that the stress intensity factor produced by a point force $F^\pm$ applied to the upper/lower crack surface at a distance $a^\pm$ 
behind the crack tip has the form:
\beq
\label{SIF_4}
K_\modIII^{(0,\pm)} = -\frac{1 \pm \eta}{\sqrt{2\pi}} F^\pm \big(a^\pm\big)^{-1/2}.
\eeq

Correspondingly, the components of the displacement gradient computed at an arbitrary point $\bY_\pm = (d\cos\phi,d\sin\phi)$ can be evaluated 
for the point force $F^+_j$ applied at the point $(-a^+_j,0+)$ as follows
\beq
\barr{l}
\ds
\left.\frac{\partial u^{(0,j)}}{\partial x_1}\right|_{\bY_\pm} = 
\frac{F^+_j}{\pi d (\mu_+ + \mu_-) (2\cos\phi + a^+_j/d + d/a^+_j)} 
\left\{ \frac{\mu_-}{\mu_\pm} 
\left( \sqrt{\frac{a^+_j}{d}}\sin\frac{\phi}{2} + \sqrt{\frac{d}{a^+_j}}\sin\frac{3\phi}{2} \right) \right. \\[6mm]
\ds
\left. \vphantom{\sqrt{\frac{a^+_j}{d}}} + \left[ \sin^2\phi - \frac{1}{2} \left( \frac{a^+_j}{d} - \frac{d}{a^+_j} \right) \cos\phi \right] 
\right\},
\earr
\eeq
\beq
\barr{l}
\ds
\left.\frac{\partial u^{(0,j)}}{\partial x_2}\right|_{\bY_\pm} = 
\frac{F^+_j}{\pi d (\mu_+ + \mu_-) (2\cos\phi + a^+_j/d + d/a^+_j)} 
\left\{ -\frac{\mu_-}{\mu_\pm}
\left( \sqrt{\frac{a^+_j}{d}}\cos\frac{\phi}{2} + \sqrt{\frac{d}{a^+_j}}\cos\frac{3\phi}{2} \right) \right. \\[6mm]
\ds
\left. \vphantom{\sqrt{\frac{a^+_j}{d}}} - \sin\phi \left[ \cos\phi + \frac{1}{2} \left( \frac{a^+_j}{d} - \frac{d}{a^+_j} \right) \right] 
\right\},
\earr
\eeq
and for the point force $F^-_j$ applied at $(-a^-_j,0-)$ we have
\beq
\barr{l}
\ds
\left.\frac{\partial u^{(0,j)}}{\partial x_1}\right|_{\bY_\pm} = 
\frac{F^-_j}{\pi d (\mu_+ + \mu_-) (2\cos\phi + a^-_j/d + d/a^-_j)} 
\left\{ \frac{\mu_+}{\mu_\pm} 
\left( \sqrt{\frac{a^-_j}{d}}\sin\frac{\phi}{2} + \sqrt{\frac{d}{a^-_j}}\sin\frac{3\phi}{2} \right) \right. \\[6mm]
\ds
\left. \vphantom{\sqrt{\frac{a^-_j}{d}}} - \left[ \sin^2\phi - \frac{1}{2} \left( \frac{a^-_j}{d} - \frac{d}{a^-_j} \right) \cos\phi \right] 
\right\},
\earr
\eeq
\beq
\barr{l}
\ds
\left.\frac{\partial u^{(0,j)}}{\partial x_2}\right|_{\bY_\pm} = 
\frac{F^-_j}{\pi d (\mu_+ + \mu_-) (2\cos\phi + a^-_j/d + d/a^-_j)} 
\left\{ -\frac{\mu_+}{\mu_\pm}
\left( \sqrt{\frac{a^-_j}{d}}\cos\frac{\phi}{2} + \sqrt{\frac{d}{a^-_j}}\cos\frac{3\phi}{2} \right) \right. \\[6mm]
\ds
\left. \vphantom{\sqrt{\frac{a^-_j}{d}}} + \sin\phi \left[ \cos\phi + \frac{1}{2} \left( \frac{a^-_j}{d} - \frac{d}{a^-_j} \right) \right] 
\right\}.
\earr
\eeq
Here the superscript $\pm$ corresponds to the surface where the point force $F^\pm_j$ is applied.

\section{Refined asymptotic formulae}
\label{sec03}

The perturbation of SIF produced by a linear defect (microcrack/rigid line inclusion) of length $2l$, centred at the point 
$\bY = (d\cos\phi,d\sin\phi)$, making an angle $\alpha$ with the positive $x_1$-direction and situated away from the interfacial crack 
tip ($l/d \ll 1$) was obtained by \cite{roaz-2011}: 
\beq
\label{SIF_accurate} 
\Delta K_\modIII = - \sqrt{\frac{2}{\pi}}
\frac{\mu_+\mu_-}{\mu_+ + \mu_-} \bB(d,\phi) \cdot \bmM(l,\alpha)
\bc(d,\phi)+O\left(\frac{l^4}{d^4}\right), 
\eeq 
where 
\beq
\bB(d,\phi) = \sum_{j}\left[ \left.\frac{\partial
u^{(0,j)}}{\partial x_1}\right|_{\bY}, \left.\frac{\partial
u^{(0,j)}}{\partial x_2}\right|_{\bY} \right], \quad 
\bc(d,\phi) = \frac{1}{2 d^{3/2}} \left[ -\sin\frac{3\phi}{2},
\cos\frac{3\phi}{2} \right], 
\eeq 
and $\bmM(l,\alpha)$ is the dipole matrix:

\noindent
-- for micro-crack 
\beq 
\bmM_{cr}(l,\alpha) = -\pi l^2 \left[
\barr{cc}
\sin^2\alpha & -\sin\alpha\cos\alpha \\[3mm]
-\sin\alpha\cos\alpha & \cos^2\alpha
\earr
\right],
\eeq
-- for rigid line inclusion 
\beq 
\bmM_{in}(l,\alpha) = \pi l^2 \left[
\barr{cc}
\cos^2\alpha & \sin\alpha\cos\alpha \\[3mm]
\sin\alpha\cos\alpha & \sin^2\alpha \earr \right]. 
\eeq 
Finally, in the presence of several micro-cracks/inclusions, formula (\ref{SIF_accurate}) will contain a sum over all defects.

Formula (\ref{SIF_accurate}) can be written in the form similar to that of (\ref{gong_1995}):
\begin{equation}
\label{MMMP_final} 
\frac{\Delta K_\modIII}{K_\modIII^{(0)}} = \frac{l^2}{d^2}\frac{\mu_ + \mu_-}{\mu_+ + \mu_-}G(\phi,\alpha) + O\left(\frac{l^4}{d^4}\right),
\end{equation}
where the function $G$ is different for the micro-crack and the rigid line inclusion:
\begin{equation}
\label{G_cr} 
G_{cr} = \cos\left(\frac{3\phi}{2} - \alpha\right)
\frac{\sqrt{\pi d}}{\sqrt{2}K_\modIII^{(0)}}\bB(d,\phi) \cdot [-\sin\alpha,\cos\alpha],
\end{equation}
\begin{equation}
\label{G_in} 
G_{in} = \sin\left(\frac{3\phi}{2} - \alpha\right)
\frac{\sqrt{\pi d}}{\sqrt{2}K_\modIII^{(0)}}\bB(d,\phi) \cdot [\cos\alpha,\sin\alpha].
\end{equation}
We would like to note that in the case when $\alpha-3\phi/2 = \pm\pi/2$ respective formulae (\ref{gong_1995}) as well as (\ref{MMMP_final}) 
and (\ref{G_cr}) provide the same (zero) results regardless of particular values of all other parameters. The same is valid for the rigid 
line inclusion when $3\phi/2-\alpha = 0$ or $\pm\pi$.

In what follows, we compare the numerical findings and asymptotic approximations and comment on applicability of the results of 
\cite{gong-1995}.

First, we compare numerically formula (\ref{gong_1995}) with (\ref{MMMP_final}) in the case when the main crack lies in a homogeneous plane 
($\mu_+ = \mu_- = \mu$) and a microcrack of the length $l = 0.01$ is situated at the distance $d = 1$ from the crack tip (that is, 
$\varepsilon = 0.01$). A symmetrical loading consisting of two point forces $F^+ = F^- = F$ acting at a distance $a$ from the crack tip is 
applied.

\begin{figure}[!htcb]
\begin{center}
\includegraphics[width=7.5cm]{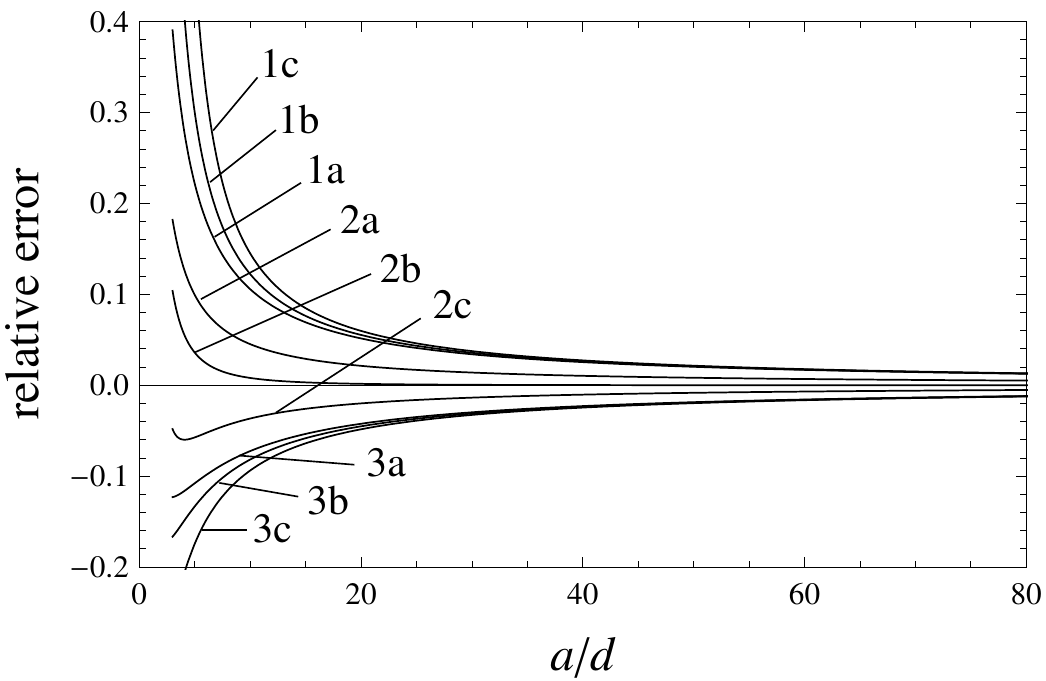}
\caption{\footnotesize Relative error of the formula (\ref{gong_1995}) in comparison with more accurate formula (\ref{MMMP_final}) as a 
function of the dimensionless parameter $a/d$ in the case of a homogeneous plane with the main crack and a micro-crack situated at a distance 
$d$ from the crack tip. A symmetrical point loading is prescribed on the crack surfaces at a distance $a$. Other parameters $\alpha$ and $\phi$ 
describe the position and orientation of the micro-crack (see Fig.\ \ref{fig01}). The following values were used in the computations: 
$\alpha = 0;\pi/3.9;\pi/2$ (cases 1,2,3), $\phi = \pi/4,\pi/2,3\pi/4$ (cases a, b, c respectively).}
\label{fig02}
\end{center}
\end{figure}

In Fig.\ \ref{fig02} we present the relative error for the case when one applies (\ref{gong_1995}) instead of (\ref{MMMP_final}). Other parameters 
involved in the computations are: $\alpha = 0;\pi/3.9;\pi/2$ (cases 1,2,3 respectively), $\phi = \pi/4,\pi/2,3\pi/4$ (cases a, b, c respectively).

One can see there is a pronounced contribution from the load applied close to the crack tip. Only if the ratio $a/d$ is big enough the 
asymptotic approximation (\ref{gong_1995}) is sufficiently accurate (for example, for $a/d>100$ the relative error is less than $1\%$).

For a separate point force applied on the upper ($F^+$) or lower ($F^-$) crack surface we have the following asymptotic formula (with 
the accuracy $O\left(\left(d/a\right)^{3/2} \right)$
$$
\left.\frac{\partial u^{(0)}}{\partial x_1}\right|_{\bY_\pm} = 
\frac{F^+}{\pi d(\mu_+ + \mu_-)}
\left\{-\frac{1}{2}\cos\phi + \frac{\mu_-}{\mu_\pm}\sqrt{\frac{d}{a^+}}\sin\frac{\phi}{2}+\frac{d}{a^+}\right\},
$$
$$
\left.\frac{\partial u^{(0)}}{\partial x_1}\right|_{\bY_\pm} = 
\frac{F^-}{\pi d(\mu_+ + \mu_-)}
\left\{\frac{1}{2}\cos\phi + \frac{\mu_+}{\mu_\pm}\sqrt{\frac{d}{a^-}}\sin\frac{\phi}{2}-\frac{d}{a^-}\right\},
$$
$$
\left.\frac{\partial u^{(0)}}{\partial x_2}\right|_{\bY_\pm} = 
\frac{F^+}{2\pi d(\mu_+ + \mu_-)}
\left\{-\sin\phi - \frac{2\mu_-}{\mu_\pm}\sqrt{\frac{d}{a^+}}\cos\frac{\phi}{2}\right\},
$$
\begin{equation}
\label{asymptotics} 
\left.\frac{\partial u^{(0)}}{\partial x_2}\right|_{\bY_\pm} = 
\frac{F^-}{2\pi d(\mu_+ + \mu_-)}
\left\{\sin\phi - \frac{2\mu_+}{\mu_\pm}\sqrt{\frac{d}{a^-}}\cos\frac{\phi}{2}\right\}.
\end{equation}

Using these representations we can perform asymptotic analysis for an arbitrary number of points forces to obtain the final result for 
SIFs in the case of a single defect situated at the point $\bY_\pm$:

\noindent
-- for a micro-crack 
\begin{equation}
\label{gong_crack} 
\frac{\Delta K_\modIII}{K_\modIII^{(0)}}=
\frac{1}{2}\frac{\mu_\mp}{\mu_+ +
\mu_-}\frac{l^2}{d^2}\left(\cos\left(\frac{3\phi}{2}-\alpha\right)
\cos\left(\frac{\phi}{2}-\alpha\right)+\chi_1\right),
\end{equation}
-- for a rigid line inclusion
\begin{equation}
\label{gong_rigid} 
\frac{\Delta K_\modIII}{K_\modIII^{(0)}}=
\frac{1}{2}\frac{\mu_\mp}{\mu_+ +
\mu_-}\frac{l^2}{d^2}\left(\sin\left(\frac{3\phi}{2}-\alpha\right)
\sin\left(\frac{\phi}{2}-\alpha\right)+\chi_2\right).
\end{equation}
The terms $\chi_1,\chi_2$ can be estimated as $O\left(\max\{d,b\}/a \right)$. This result shows that the second term of the order $O(l^4/d^4)$ 
in the formula (\ref{gong_1995}) is significant if the following estimate is valid $\max\{d,b\}/a = o(l^2/d^2)$, otherwise there are other 
essential terms in the asymptotics larger than $O(l^4/d^4)$.

Note that, when computing (\ref{gong_crack}) and (\ref{gong_rigid}), the first terms in (\ref{asymptotics}) have disappeared after summation 
as the load is self-balanced.

To give an additional comparison with (\ref{gong_crack}), we also consider a special case when the crack faces are loaded by a ``three-point'' 
loading system consisting of a point force $F$ acting on the upper crack face at a distance $a$ behind the crack tip and two point forces $F/2$ 
acting on the lower crack face at distances $a - b$ and $a + b$ behind the crack tip. The parameter $b$ is fixed. We  assume here that 
$b/a = o(d/a)$ and obtain
\beq
\frac{\Delta K_\modIII}{K_\modIII^{(0)}} = \frac{1}{2}
\frac{l^2}{d^2} \frac{\mu_\mp}{\mu_+ + \mu_-} \cos\left(\frac{3\phi}{2} - \alpha\right)
\left\{\cos\left(\frac{\phi}{2} - \alpha\right) + \frac{d}{a} \cos\left(\frac{\phi}{2} + \alpha\right) +
O\left(\frac{b^2}{a^2}\right)\right\} + O\left(\frac{l^4}{d^4}\right),
\eeq
for a microcrack, while for a rigid line inclusion it takes the form: 
\beq 
\frac{\Delta K_\modIII}{K_\modIII^{(0)}} = \frac{1}{2} 
\frac{l^2}{d^2} \frac{\mu_\mp}{\mu_+ + \mu_-} \sin\left(\frac{3\phi}{2} - \alpha\right)
\left\{\sin\left(\frac{\phi}{2} - \alpha\right) - \frac{d}{a} \sin\left(\frac{\phi}{2} + \alpha\right) +
O\left(\frac{b^2}{a^2}\right)\right\} + O\left(\frac{l^4}{d^4}\right),
\eeq

\section{Numerical computations}
\label{sec04}

Before the discussion of computations we would like to point out that the simplified asymptotic formulae (\ref{gong_crack}) and 
(\ref{gong_rigid}) give similar predictions regardless of the bimaterial parameter $\eta$ (the ratio of the shear moduli $\mu_+/\mu_-$). 
Of course, the value of the stress intensity factors will be a function of this ratio. The more accurate formula (\ref{MMMP_final}) 
indicates the influence of possible non-symmetry of the applied load.

\begin{figure}
\centering
\subfigure[micro-crack]{\includegraphics[width=6cm]{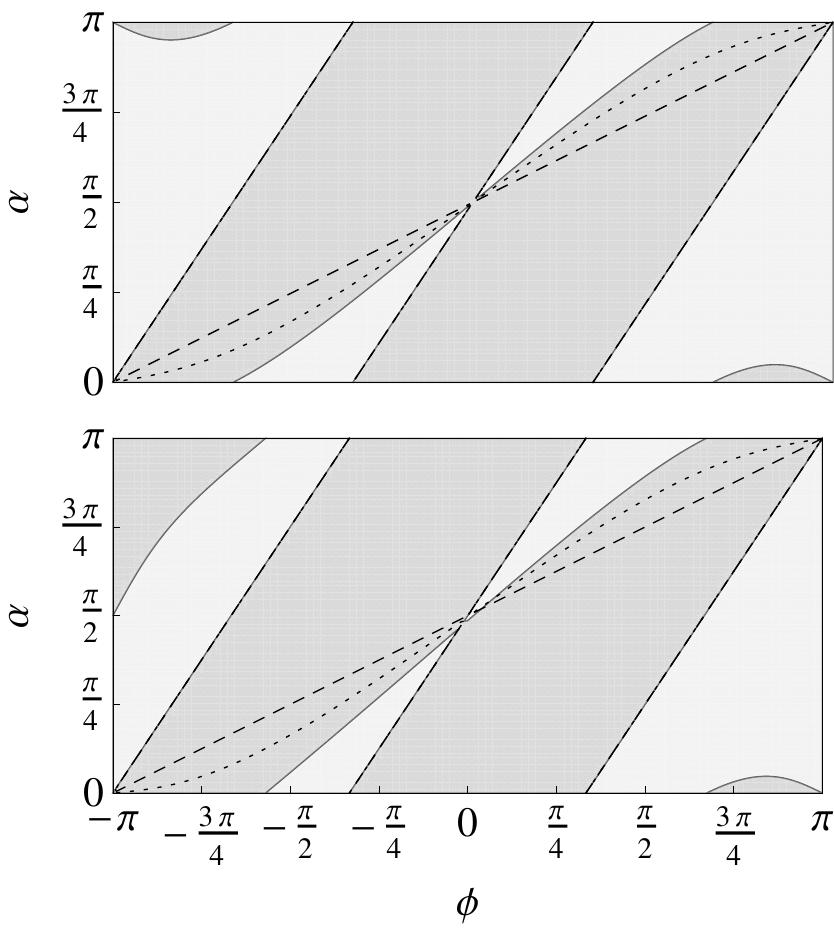}}\qquad
\subfigure[rigid line inclusion]{\includegraphics[width=6cm]{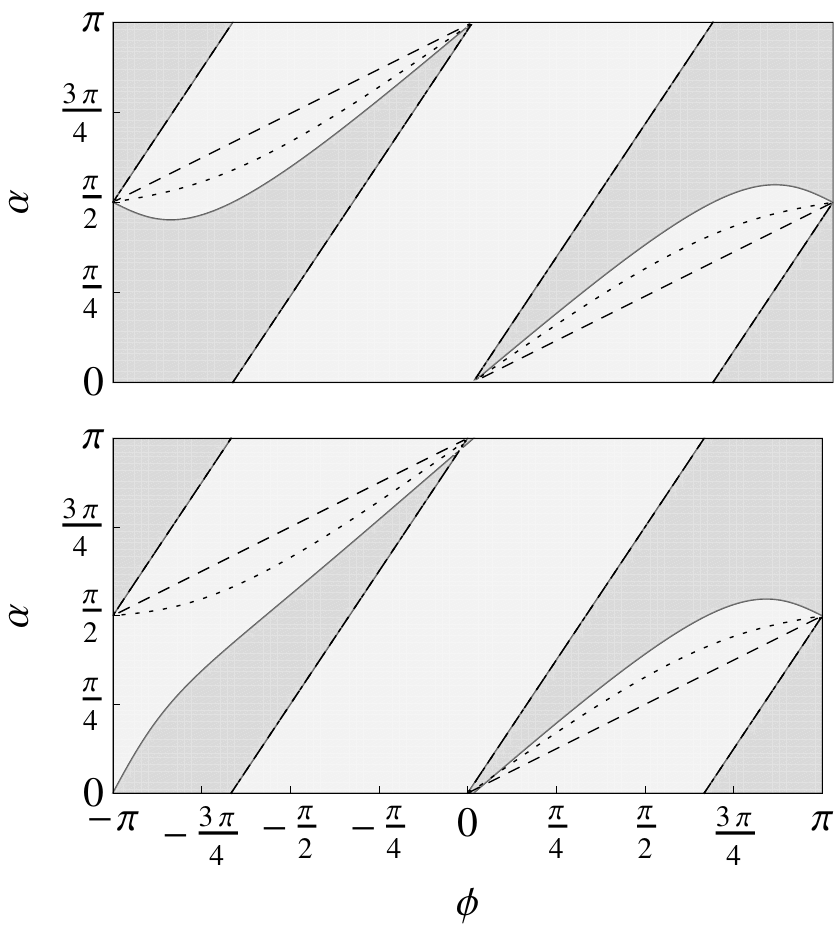}}
\caption{\footnotesize `` Shielding'' (light grey) and ``amplification'' (dark grey) regions created by a micro-crack (a) or a rigid line 
inclusion (b), whose position relative to the $x_1$-axis is characterised by the angles $\phi$ and $\alpha$. Upper part: two-point loading 
($b = 0$) at a distance $a$ from the crack tip. Lower part: non-symmetrical three-point loading ($b = 1$) at a distance $a$ from the crack tip 
(see Fig.\ \ref{fig01}(a)). The borders between the regions for $a = 2$, $a = 4$ and Gong's approximation are shown by solid, dotted and 
dashed lines, respectively.}
\label{fig0304}
\end{figure}

We first present the results for a homogeneous body ($\mu_+ = \mu_- = \mu$), see Fig.\ \ref{fig0304}. In the computations, we use the value of 
the small parameter $\varepsilon = 0.01$ which describes a relative size of the small defect in comparison with the distance to the crack tip 
($d = 1$ and $l = 0.01$).

On the diagrams, the horizontal axis stands for the angle $\phi$ ($\phi \in (-\pi,\pi)$, see Fig.\ \ref{fig01}) defining the position of the 
center of the micro-defect with respect to the crack tip. On the vertical axis we measure the value of the angle $\alpha$ ($0 < \alpha < \pi$). 
The diagrams show the borders between the region where $\Delta K < 0$ ({\it shielding} effect) and $\Delta K > 0$ ({\it amplification} of the SIF 
due to the presence of the small defect). The respective regions are shadowed by light gray ({\it shielding}) and dark gray ({\it amplification}).

Fig.\ \ref{fig0304}(a)-upper corresponds to a two-point (symmetrical) load situated at a distance $a$ from the crack tip in the case of a small 
micro-crack placed at the distance $d$ from the crack tip. As it is clear from the presented results, the distance between the load support and 
the crack tip plays an important role (solid, dotted and dashed lines are different in the middle of the diagram). Moreover, in the case $a = 2$ 
when the load is close enough to the crack tip (but still $a > d$) new small {\it amplification} regions appear in the corners of the picture. 
It is natural that the diagram is completely symmetrical with respect to the angle $\phi$ as the load is symmetrical and the material parameters 
do not influence the diagram.

The results change when we consider a three-point non-symmetrical load (as shown in Fig.\ \ref{fig01}(a)). In Fig.\ \ref{fig0304}(a)-lower 
we set the parameter $b$, the distance between two point forces on the lower crack surface, to 1.

In Fig.\ \ref{fig0304}(b)-upper we consider the same symmetrical load as in Fig.\ \ref{fig0304}(a)-upper, however now a small defect is 
represented by the rigid line inclusion placed at the same distance $d$ from the crack tip. The {\it shielding} and {\it amplification} regions 
are completely different in comparison with the micro-crack case. As in Fig.\ \ref{fig0304}(a)-upper, due to the symmetry of the load, the 
diagram of Fig.\ \ref{fig0304}(b)-upper is symmetric with respect to the angle $\phi$.

In Fig.\ \ref{fig0304}(b)-lower we consider a three-point non-symmetrical load as in Fig.\ \ref{fig0304}(a)-lower, but we replace the 
micro-crack by a rigid line inclusion. As in the previous cases, when $a$ is large (and even for $a = 4$) non-symmetry in the load is hard 
to observe. However, in the case $a = 2$ the diagram shows a clear asymmetry.

\begin{figure}
\centering
\subfigure[micro-crack]{\includegraphics[width=6cm]{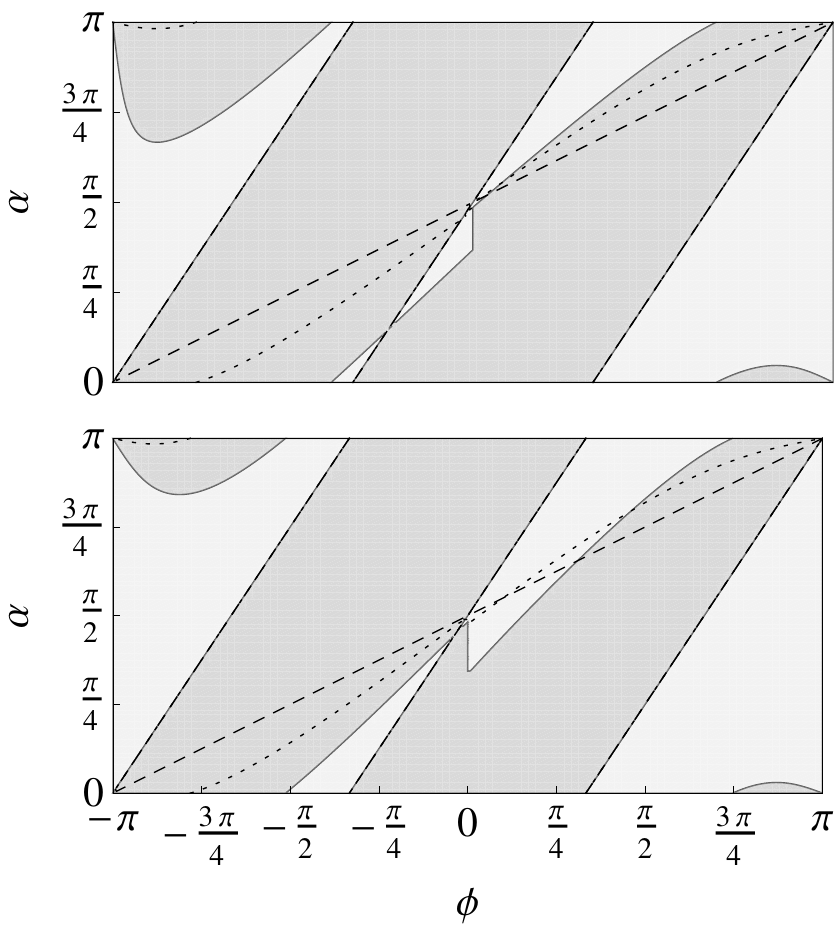}}\qquad
\subfigure[rigid line inclusion]{\includegraphics[width=6cm]{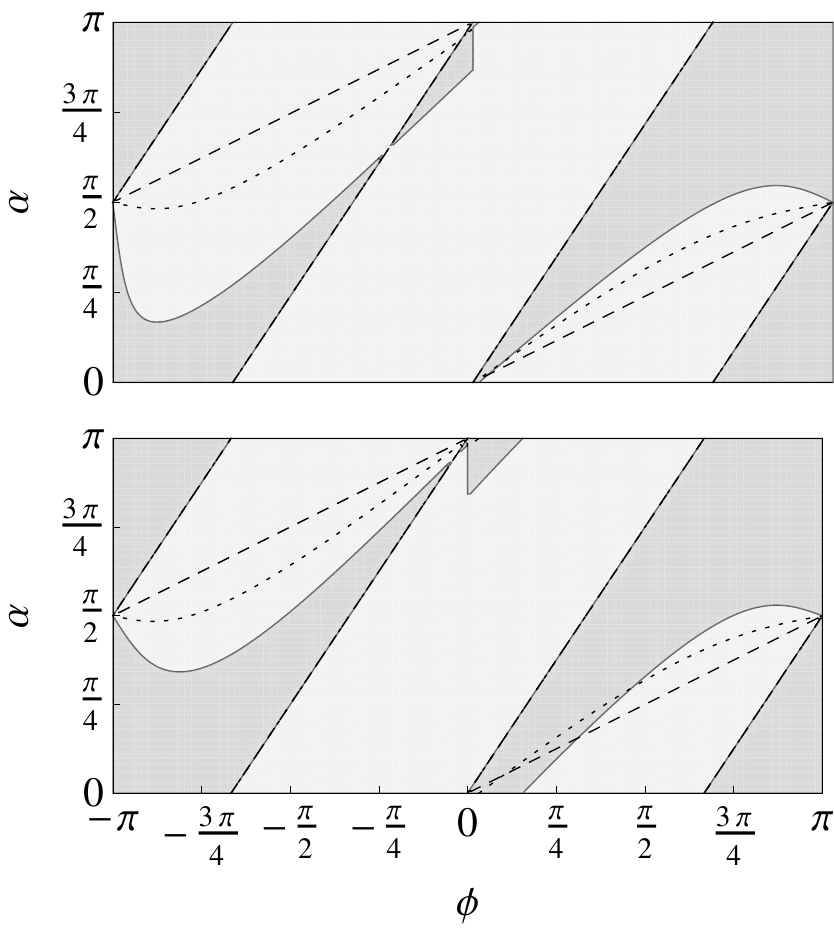}}
\caption{\footnotesize ``Shielding'' (light grey) and ``amplification'' (dark grey) regions created by a micro-crack (a) or a rigid line 
inclusion (b), whose position is characterised by the angles $\phi$ and $\alpha$ (see Fig.\ \ref{fig01}). The diagrams are for a 
three-point loading with $b = 1.9$, and the material parameters ratio $\mu_+/\mu_- = 0.1$ (upper part) and $\mu_+/\mu_- = 10$ (lower part). 
The borders between the regions for $a = 2$, $a = 4$ and $a = 10000$ are shown by solid, dotted and dashed lines, respectively.}
\label{fig0506}
\end{figure}

Finally we would like to discuss the influence of the inhomogeneity on the shielding effect. As we just mentioned, it can be only observable 
for a highly pronounced non-symmetrical loading which is applied close to the crack tip. For this reason in Fig.\ \ref{fig0506} we consider the 
aforementioned three-point loading for the case when $b = 1.9$ and $a = 2, 4$ and $10000$. Two ratios of the material parameters were used in 
the computations: $\mu_+/\mu_- = 0.1$ (upper part) and $\mu_+/\mu_- = 10$ (lower part).

The diagrams of Fig.\ \ref{fig0506}(a) correspond to a micro-crack defect, while those of Fig.\ \ref{fig0506}(b) correspond to a rigid line 
inclusion.

\section{Discussion and conclusions}
\label{sec05}

In this paper we have discussed asymptotic formulae for the perturbation of the Mode III SIF due to the presence of a small defect at a 
large distance from the tip of an interfacial crack ($\varepsilon = l/d \ll 1$). We have derived a novel accurate 
{\em analytical} formula based on the dipole matrix and weight-function approach. It was noted that the position of a small defect influences 
the SIF leading to {\it shielding} and {\it amplification} effects for various geometrical and material parameters. We have also shown that, 
for a Mode III deformation, the material inhomogeneity (dissimilar plane) plays an important role only in the case of a non-symmetric loading 
applied close to the crack tip.

The method exploited here allows to define Mode III SIF for an arbitrary load by generalising the asymptotic representation (\ref{dipolar}) 
and to estimate the error of the asymptotic approximation. Finally, we would like to emphasise that, due to the linearity of the problem, the 
influence of a finite number of small defects can be computed by a simple summation of the formulae (\ref{MMMP_final}) over every particular defect. 
The dipole matrices for clusters of small defects would have to be employed if the number of micro-cracks/inclusions becomes large.

Compared to the earlier publications on related topics (e.g. \cite{petrova-2000}, \cite{Wang1}, \cite{Wang2}, \cite{Bo}), our results allow for an 
explicit analytical representation of the asymptotic solution. Also, the approach based on the analysis of dipole fields is generic, and it is not 
restricted to a particular shape of defects, like micro-cracks, for example. The method is applicable equally well to the case of small voids or 
inclusions of arbitrary shapes, with smooth or non-smooth boundaries, including clouds of defects of different shapes. 

\vspace{6mm}
{\bf Acknowledgements}. This research was supported by the Research-In-Groups (RiGs) programme of the International Centre for Mathematical 
Sciences, Edinburgh, Scotland. In addition, A.P. gratefully acknowledges the support from the European Union Seventh Framework Programme 
under contract number PIEF-GA-2009-252857.

\bibliographystyle{abbrv}
\bibliography{biblio}

\end{document}